\documentstyle[11pt,newpasp,twoside]{article}
\def\edcomment#1{\iffalse\marginpar{\raggedright\sl#1\/}\else\relax\fi}
\reversemarginpar

\begin{document}
\title{Metals in the Universe and Diffuse Background Radiation}  
 \author{B E J Pagel}
\affil{Astronomy Centre, University of Sussex, Brighton BN1 9QJ, UK} 

\begin{abstract}
An attempt is made to guess the overall cosmic abundance of `metals' and 
the contribution made by the energy released in their production to the 
total intensity of extragalactic background light (EBL). With a comparable 
or somewhat larger amount coming from white dwarfs, and a probably quite 
modest contribution from AGNs, one can fairly easily account for the lower 
end of the range of existing estimates for the total EBL intensity (50 to 60 
nwt m$^{-2}$ sterad$^{-1}$), but it seems more difficult should some 
higher estimates (90 to 100 in the same units) prove to be correct.   
\end{abstract}

\section{Introduction} 
There are certain more or less well or badly determined integral constraints
on the past history of star formation in the universe. 
\footnotetext{Invited talk to Vulcano Workshop: {\em Chemical Enrichment of 
Intracluster and Intergalactic Medium}, May 14--18 2001, F. Matteucci \& 
F. Giovannelli (eds.), ASP Conference Series.} 
These include 
\begin{itemize} 
\item Cosmic baryon density $\Omega_{\rm B}$, now fairly well determined 
both from primordial deuterium (O'Meara et al 2000) and from the CMB fluctuation 
spectrum (e.g. Turner 2001). 
\item Cosmic mass density of stars $\Omega_ *$, rather less well determined 
as it 
involves a combination of luminosity-density measurements with an IMF 
and evolutionary population synthesis models.  
\item Extragalactic background light (EBL) intensity, now known within a 
factor of 2 or so from COBE (FIRAS and DIRBE) and galaxy counts in the 
optical and near IR (e.g. Gispert, Lagache \& Puget 2000).      
\item Cosmic abundance of `metals', $\Omega_{\rm Z}$, due to the 
heavy-element content of stars, the interstellar medium and the 
intergalactic medium, a quantity that is very poorly known and largely 
a matter of guesswork.  In this talk I shall nevertheless make some guesses, 
so that at least one can see more easily how things relate to one another. 
In particular, the metallicity $Z_{\rm IGM}$ of the intergalactic 
medium has tended to be either neglected or underestimated in models of 
the past star formation 
rate, and it is of interest to ask about its relation with EBL.    

\end{itemize} 
\newpage 
\section{The cosmic inventory} 

\begin{table} 
\begin{center} 
\caption{Inventory of cosmic baryons and `metals'}  
\vspace{5mm} 

Densities expressed as $\Omega$, in units of $\rho_{{\rm crit}} 
= 1.54\times 10^{11}h_{70}^{2}\,M_{\odot}\,{\rm Mpc}^{-3}$

\begin{tabular}{ll}

\hline
\\ 

All baryons from BBNS\\  
(D/H = $3\times 10^{-5}$ $^ a$)& $0.04\,h_{70}^{-2}$ \\ \hline  
\\ 
Stars in spheroids &      $0.0026\,h_{70}^{-1}$ $^ b$
\\ \\  
Stars in disks &    $0.0009\,h_{70}^{-1}$ $^ b$ 
\\ \\  
Total stars &      $0.0035\,h_{70}^{-1}\, ^ b$   
\\ \\  
Cluster hot gas     &   $0.0026\,h_{70}^{-1.5}$ $^ b$
\\ \\  
Group/field hot gas &      $0.014\,h_{70}^{-1.5}$ $^ b$ ($0.004h_{75}^{-1}$
in O {\sc vi} systems $^ c$)  
\\ \\ 
Total stars + gas &   $0.021\,h_{70}^{-1.5}$ $^ b$ \\ \\  
Machos + LSB gals& ?? $ ^b$ \\ 
  \hline \\   
$\Omega_{{\rm Z}}$ (stars, $Z = 0.02\;^ d$) 
&   $7\times 10^{-5}\,h_{70}^{-1}$   \\
\\ 
$\Omega_{{\rm Z}}$ (hot gas, $Z=.006$)   
&   $1.0\times 10^{-4}\,h_{70}^{-1.5}$ $^ b$\\
  & $1.2\times 10^{-4}\,h_{70}^{-1.3}$ $^ e$\\ 
\\ 
Yield $\rho_{{\rm Z}}/\rho_{*}$ &  $0.051\,h_{70}^{-0.3}\;\;(\simeq 
3Z_{\odot}$!)\\   
\hline \\  
Damped Ly-$\alpha$ (H {\sc i})&   $0.0015\,h_{70}^{-1}$ $^{b,f}$\\
\\ 
Ly-$\alpha$ forest (H$^ +$)&   $0.04\,h_{70}^{-2}$ $^{b,g}$\\ 
 \hline \\ 
Gals + DM halos\\ 
(M/L = $210\,h_{70}$)&   0.25 $^{b,h}$\\ \\
All matter \\ 
($f_{{\rm B}}= .056\,h^{-1.5}$)   & $0.37\, h_{70}^{-0.5}$ $^{b,i}$  \\ 
\hline \tableline  
\end{tabular} 
\end{center} 
{\small 
$^ a$ O'Meara et al 200l; but see also Pettini \& Bowen 2001;   
$^ b$ Fukugita, Hogan \& Peebles 1998; 
$^ c$ Tripp, Savage \& Jenkins 2000;  
$^ d$ Edmunds \& Phillipps 1997; 
$^ e$ Mushotzky \& Loewenstein 1997;  
$^ f$ Storrie-Lombardi, Irwin \& MacMahon 1996; 
$^ g$ Rauch, Miralda-Escud\' e, Sargent et al 1998;
$^ h$ Bahcall, Lubin \& Dorman 1995;  
$^ i$ White \& Fabian 1995.  
}
\end {table} 
 
A useful starting point is the cosmic baryon budget drawn up by 
Fukugita, Hogan \& Peebles (1998), hereinafter FHP, shown in the 
accompanying table. The total from 
Big Bang nucleosynthesis (BBNS) adopted here agrees quite well with 
the amount of intergalactic gas at a red-shift of 2 to 3 deduced from 
the Lyman forest, but exceeds the present-day stellar (plus cold gas) 
density by an order of magnitude. \footnote{The stellar density taken here from 
FHP is based on B-luminosity density estimates and might be revised 
upwards by 50 per cent in light of SDSS commissioning data (Blanton et al 
2000)  or downwards by 20 per cent in light of 2dF red-shifts plus 
2MASS K-magnitudes (Cole et al 2000); in either case we are following 
FHP in assuming the IMF by Gould, Bahcall \& Flynn (1996), which has 0.7 
times the $M/L_{\rm V}$ ratio for old stellar populations compared to 
a Salpeter function with lower cutoff at $0.15M_{\odot}$.}   

FHP pointed out that a dominant and uncertain contribution to the 
baryon budget comes from intergalactic ionized gas, not readily 
detectable because of its high temperature and low density. The number 
which I quote is based on the assumption that the spheroid 
star-to-gravitational mass ratio and baryon fraction are the same in clusters 
and the field, an assumption that had also been used previously by 
Mushotzky \& Loewenstein (1997). The resulting total star-plus-gas density 
is within spitting distance of $\Omega_{\rm B}$ from BBNS, but leaves a 
significant-looking shortfall which may be made up by some combination 
of MACHOs and low surface-brightness galaxies; it is not clear that a 
significant contribution from the latter has been ruled out (cf O'Neil
2000).        

\section{Global abundances and yields} 

We now have the tricky task of estimating the total heavy-element content 
of the universe. Considering stars alone, it seems reasonable to adopt 
solar $Z$ as an average, but the total may be dominated by the still unseen 
intergalactic gas, which Mushotzky \& Loewenstein argue to have the same 
composition as the hotter, denser gas seen in clusters of galaxies, i.e. 
about 1/3 solar.\footnote{This refers to iron abundance, the relation of 
which to the more energetically relevant quantity $Z$ is open to some  
doubt.  Papers given at this conference indicate an SNIa-type mixture in 
the immediate surroundings of cD galaxies with maybe a more SNII-like 
mixture in the intra-cluster medium in general; for simplicity I assume 
the mixture to be solar.} It could be the case, though, that the metallicity 
of the IGM is substantially lower in light of the metallicity-density 
relation predicted by Cen \& Ostriker (1999) and in that of the low 
metallicities found in low red-shift Ly-$\alpha$ clouds by Shull et al (1998). 
Against this, we have 
neglected any metals contained in LSB galaxies or whatever makes up the 
shortfall between $\Omega_{\rm IGM}$ and $\Omega_{\rm B}$, so we are being 
conservative in our estimate of $\Omega_{Z}$. 

\begin{figure*}[htbp] 
\vspace{4.5cm} 
\includegraphics{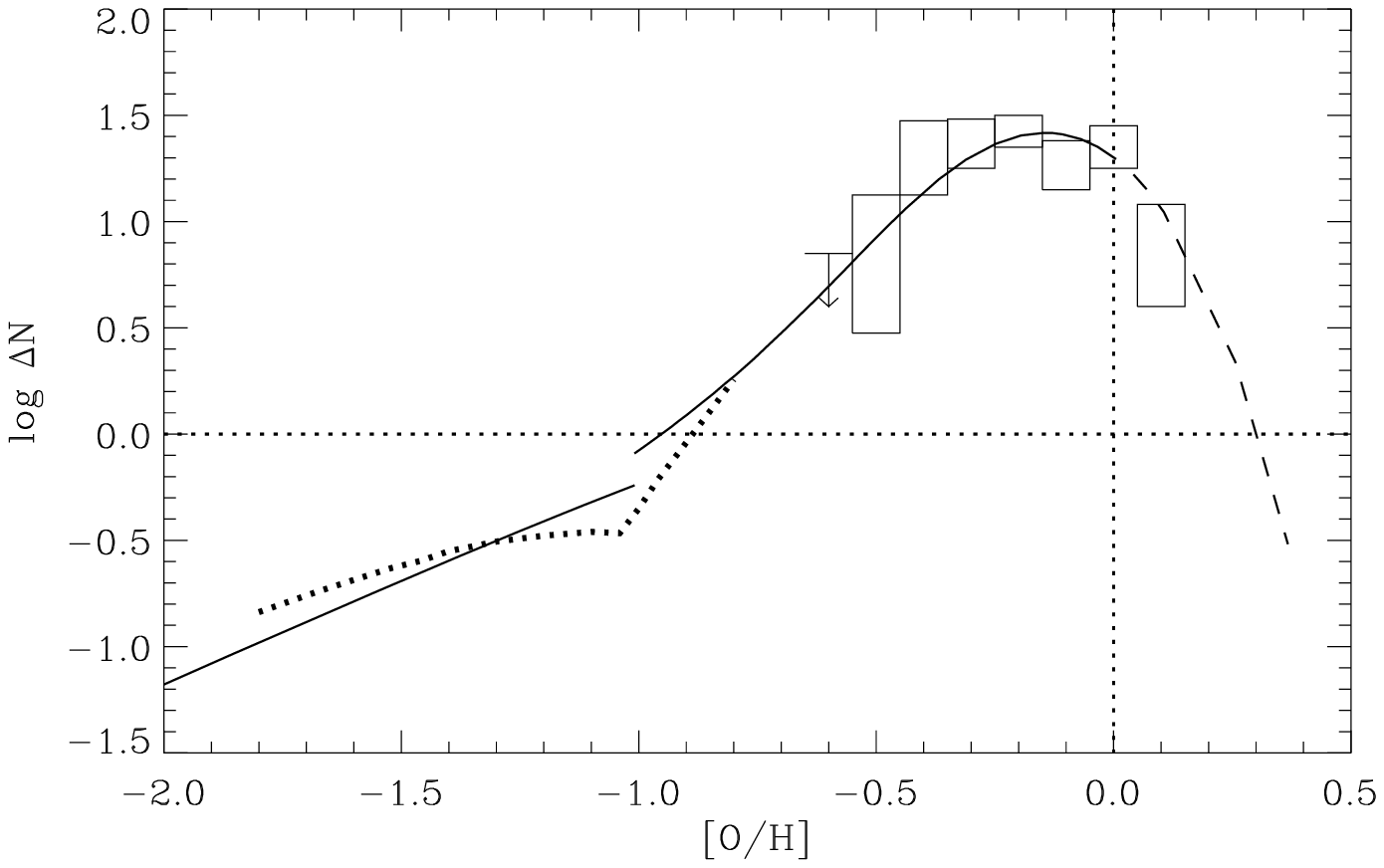} 
\caption{Oxygen abundance distribution function in the solar neighbourhood, 
after Pagel \& Tautvai\v sien\. e (1995).} 
\end{figure*} 

The mass of heavy elements in the universe is related to that of stars 
through the yield, defined as the mass of `metals' synthesised and 
ejected by a generation of stars divided by the mass left in form of 
long-lived stars or compact remnants (Searle \& Sargent 1972). The        
yield may be predicted by a combination of an IMF with models of stellar 
yields as a function of mass, or deduced empirically by applying a 
galactic chemical evolution (GCE) model to a particular region like the 
solar neighbourhood and comparing with abundance data.  E.g. Fig 1 shows 
an abundance distribution function for the solar neighbourhood plotted in 
a form where in generic GCE models the maximum of the curve gives the 
yield directly, and it is a bit below $Z_{\odot}$.  Similar values are 
predicted theoretically using fairly steep IMFs like that of Scalo 
(1986).  In Table 1, on the other hand, if we divide the mass of metals 
by the mass of stars, we get a substantially higher value, corresponding 
to a more top-heavy IMF. 

There are two other indications for a top-heavy IMF, one local and one 
in clusters of galaxies themselves. The local one is an investigation by 
Scalo (1998) of open clusters in the Milky Way and the LMC, where he plots 
the IMF slopes found as a function of stellar mass. The scatter is large, 
but on average he finds a Salpeter slope above $0.7M_{\odot}$ and a 
virtually flat relation (in the sense $dN/d\log m\simeq 0$) below, 
which could quite easily account for the sort of yield found in Table 1. 
The other indication is just the converse of the argument we have already 
used in guessing the abundance in the IGM: the mass of iron in the 
intra-cluster gas is found (Arnaud et al 1992) to be  
\begin{equation} 
M_{\rm Fe}/L_{\rm V} = 0.018 M_{\odot}/L_{\odot}, 
\end{equation} 
where $L_{\rm V}$ is the luminosity of E and S0 galaxies in the cluster. 
As has been pointed out by Renzini et al (1993) and Pagel (1997), given a 
mass:light ratio less than 10, we then have 
\begin{eqnarray} 
M_ {\rm Fe} ({\rm gas})/M_ * \geq 1.8\times 10^{-3} &=&1.6Z_{\odot}({\rm Fe})
\\
M_{\rm Fe}(*)/M_ * &\simeq & Z_{\odot}({\rm Fe})\\  
{\rm Yield} = M_{\rm Fe}({\rm total})/M_ *& \geq & 2.6Z_{\odot}. 
\end{eqnarray}        

The argument is very simple; the issue is just whether such high 
yields are universal or confined to elliptical galaxies in clusters. 

\section{Cosmic star formation and chemical evolution: GCE vs HDF} 

The deduction of past star formation rates from rest-frame UV radiation 
in the Hubble and other deep fields as a function of red-shift is tied 
to `metal' production through the Lilly-Cowie theorem (Lilly \& Cowie 
1987): 
\begin{eqnarray}
\rho_{\rm L}(z) & = & \nu_{\rm H}\rho_{\nu\;\rm{UV}}(z) = \frac{1} 
{\epsilon} L_{\rm{FIR}}(z) \\
& = & 0.007c^ 2 \dot{\rho}_{\rm Z}(z)(1+a)\beta^{-1} \\ 
& = & 0.018c^ 2\dot{\rho}_{\rm Z} (\rm{conv.}), 
\end{eqnarray} 
where $(1+a)\simeq 2.6$ is a correction factor to allow for production 
of helium as well as conventional metals and $\beta$ (probably between 
about 1/2 and 1) allows for nucleosynthesis products falling back into 
black-hole remnants from the higher-mass stars. $\epsilon$ is the fraction 
of total energy output absorbed and re-radiated by dust and $\nu_{\rm H}$ 
is the frequency at the Lyman limit (assuming a flat spectrum at lower 
frequencies). The advantage of this formulation is that the relationship 
is fairly insensitive to details of the IMF.

\begin{figure*}[htbp]   
\vspace{6cm} 
\includegraphics{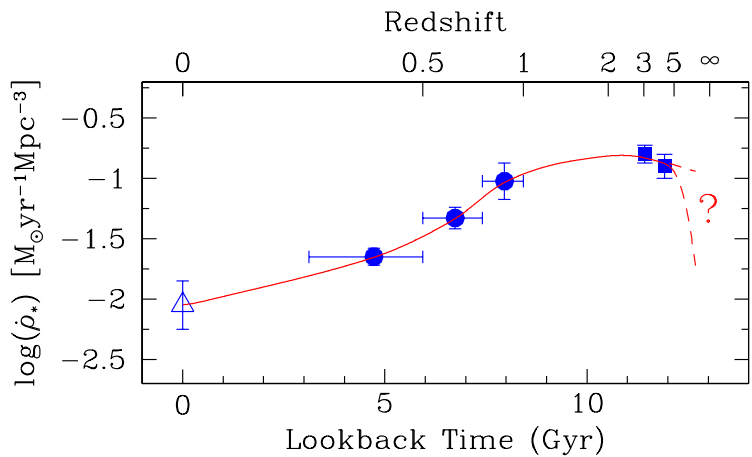}
\caption{Global comoving star formation rate density vs. lookback time 
compiled from wide-angle ground-based surveys (Steidel et al. 1999 and 
references therein) assuming E--de S cosmology with $h=0.5$, after 
Pettini (1999). Courtesy Max Pettini.}     
\end{figure*} 

Eq (7) is the same as eq (13) of Madau et al (1996), so I refer to the 
metal-growth rate derived in this way as $\dot{\rho}_{\rm Z}$(conventional).    

Assuming a Salpeter IMF from $0.1$ to $100M_{\odot}$ with all stars 
above 10$M_{\odot}$ expelling their synthetic products in SN explosions, 
one then derives a conventional SFR density through multiplication with 
the magic number 42: 
\begin{equation} 
\dot{\rho}_ *(\rm{conv.}) =42\dot{\rho}_{\rm Z}(\rm{conv.})=42\rho_{\rm L}/ 
.018c^ 2. 
\end{equation} 
In general, we shall have 
\begin{equation} 
\dot{\rho}_ *(\rm{true}) = \gamma \dot{\rho}_ *(\rm{conv.}), 
\end{equation} 
where $\gamma$ is some factor.  E.g., for the IMF adopted by FHP, 
$\gamma=0.67$, whereas for the Kroupa-Scalo one (Kroupa et al 1993) 
$\gamma=2.5$. 
 
Finally, the present stellar density is derived by integrating over the 
past SFR  and allowing for stellar mass loss in the meantime, and the 
metal density is related to this through the yield, $p$: 
\begin{eqnarray} 
\rho_ *(0,\rm{true}) &=& \alpha\gamma\int\dot{\rho}_ *(\rm{conv.})dt\\ 
\rho_{\rm Z}(0,\rm{true}) &=& p\rho_ *(0{,\rm{true}})
= \left(\frac{2.6}{1+a}\right)\,\frac{\beta}{42}\int\dot\rho_ *(\rm{conv.})dt 
\end{eqnarray}
(where $\alpha$ is the lockup fraction), whence (if $a=1.6$)  
\begin{equation} 
p = \frac{\beta}{42\alpha\gamma},  
\end{equation}   
which can be compared with $Z_{\odot}\simeq$ 1/60. It was pointed out by 
Madau et al (1996) that the Salpeter slope gives a better fit to the 
present-day stellar density than one gets from the steeper one --- a 
result that is virtually independent of the low-mass cutoff if one 
assumes a power-law IMF.   

Eq (8), duly corrected for absorption, forms the basis for numerous discussions 
of the cosmic past star-formation rate or `Madau plot'.  Among the more 
plausible ones are those given by Pettini (1999)  shown in Fig 2 and by 
Rowan-Robinson (2000), which leads to similar results and is shown to 
explain the far IR data. Taking $\gamma = 0.62$ (corresponding to a 
Salpeter IMF that is flat below $0.7M_{\odot}$) rather than Pettini's value of 
0.4 (for an IMF truncated at $1M_{\odot}$), and $\alpha =0.7$, we get the 
data in the following table.

\begin{center} 
\begin{table} 
{\large 
\caption{Inventory of stars and metals at $z=0$ and $z=2.5$} 
\label{tab1} 
\begin{tabular}{|r|c|c|} 
\hline
&$z=0$ & $z=2.5$ \\ \hline 
&&\\  
$\rho_ * = \alpha\,\gamma\int\dot{\rho_ *}{\rm (conv.)}\,dt$ & 
$3.6\times 10^ 8\,  
 M_{\odot}\,{\rm Mpc}^{-3}$ & $9\times 10^ 7\,M_{\odot}\, {\rm Mpc}^{-3}$\\ 
&&\\  
$\Omega_ *= \rho_ */1.54\times 10^{11}h_{70}^2$ & $.0024h_{70}^{-2}$ & 
$6\times 10^{-4}h_{70}^{-2}$ \\ 
$\Omega_ *$(FHP 98) & $.0035h_{70}^{-1}$ &\\ 
&&\\  
$\rho_ Z=p\rho_ *=\beta\rho_ */(42\alpha\gamma)$ & $2.0\times 10^ 7\beta\,  
M_{\odot}\,{\rm Mpc}^{-3}$ & $5\times 10^ 6 \beta \,M_{\odot}\, 
{\rm Mpc}^{-3}$\\ 
$\Omega_{Z}$ (predicted) & $1.3\times 10^{-4}\beta h_{70}^{-2}$ & 
$3.2\times 10^{-5}\beta h_{70}^{-2}$\\ 
&&\\  
$\Omega_ Z$ (stars, $Z=Z_{\odot}$) & $7\times 10^{-5}h_{70}^{-1}$ &\\ 
$\Omega_ Z$ (hot gas, $Z=0.3Z_{\odot}$) & $1.0\times 10^{-4}h_{70}^{-1.5}$&\\
&$\Rightarrow 0.5\leq\beta\leq 1.3$&\\ 
&&\\   
$\Omega_ Z$ (DLA, $Z=0.07Z_{\odot}$) & & $2\times 10^{-6}h_{70}^{-1}$\\ 
$\Omega_ Z$ (Ly. forest, $Z=0.003Z_{\odot}$) & & $1\times 10^{-6}h_{70}^{-2}$
\\
$\Omega_ Z$ (Ly. break gals, $Z=0.3Z_{\odot}$) & & ? \\
$\Omega_ Z$ (hot gas) & & ?\\ \hline       

\end{tabular} } 
\end{table}
\end{center}

Table 2 indicates that the known stars are roughly accounted for by the 
history shown in Fig 2 (or by Rowan-Robinson) and the metals also if 
$\beta$ is close to unity, i.e. the full range of stellar masses expel 
their nucleosynthesis products. At the very least, $\beta$ has to be 1/2, 
to account for metals in stars alone.  The other point arising from the 
table, made by Pettini, is that at a red-shift of 2.5, 1/4 of the stars 
and metals have already been formed, but we do not know where the 
resulting metals reside. 

\section{Extragalactic background light} 

\begin{figure*}[htbp]
\vspace{5cm}
\includegraphics{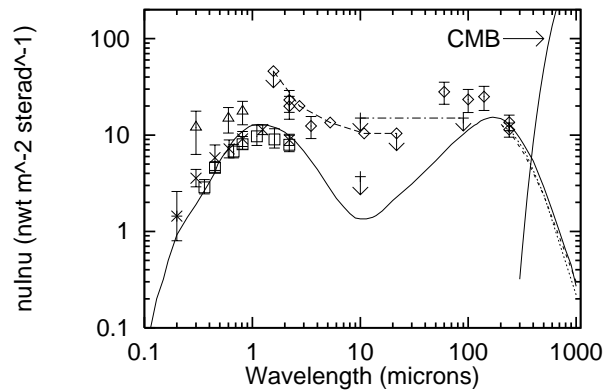}
\caption{Spectrum of extragalactic background light, based on COBE data 
after Hauser 2001 (diamonds with error bars, dotted and short-dash curves), 
Madau \& Pozzetti 
2000 (squares), Totani et al 2001 (crosses), Bernstein, Freedman \& Madore 
2001 (triangles) and Armand et al 1994 (asterisk).  The broken-line 
curve (Biller et al 1998) and horizontal dash-dot line (Hauser 2001) show  
upper limits based on lack of attenuation of high-energy 
$\gamma$-rays from AGNs and the solid curve is from the model by Pei, Fall 
\& Hauser (1999). The arrow showing an upper limit at $10\mu$m is from 
unpublished thesis work by A. Barrau, cited by Gispard, Lagache \& Puget 
(2000).}      
\end{figure*}

Fig 3 shows the spectrum of extragalactic background light (EBL) with the 
model fit by Pei, Fall \& Hauser (1999).  Gispert, Lagache \& Puget (2000) 
have estimated the total EBL $\int I_{\nu}d\nu$ based on observation to lie 
within the following limits: 

\begin{tabular}{rl} 
$\lambda\leq 6 \mu$m: & 20 to 40 nwt m$^{-2}$ sterad$^{-1}$ \\ 
$\lambda > 6 \mu$m:   & 40 to 50  \hspace{3mm}"\hspace{3mm}"\hspace{9mm} " \\
Total:                & 60 to 90 \hspace{3mm}"\hspace{3mm}"\hspace{9mm} " 
\end{tabular} 

(The total from the model of Pei, Fall \& Hauser (1999) is 55 in these units.)

We use the estimates of stellar and metal densities in Tables 1 and 2 
together with eq (7) and an assumption about the mean red-shift  of 
metal formation to derive the EBL contributions from:  
\begin{itemize}
\item Metals in stars; $Z=Z_{\odot}=0.02$:  
\begin{eqnarray} 
  \rho_ Z(*) &= &7\times 10^{-34}\,h_{70}\;\; \;{\rm gm\;cm}^{-3}\\ 
I &=& \frac{0.018c^ 3 \rho_ Z \beta^{-1}}{4\pi\langle1+z\rangle} = 
 \langle\frac{3}{1+z}\rangle\,\beta^{-1}\,\rho_ Z\times 1.3 \times 10^{34} \\ 
& \simeq & 9\beta^{-1}\,h_{70}\;\;{\rm nwt\;m}^{-2}\,{\rm sterad}^{-1}.
\end{eqnarray} 

\item Metals in diffuse gas/dark baryons, $Z\simeq 0.007$: 
\begin{eqnarray} 
\rho_ Z = \Omega_ Z \rho_{\rm crit} &=&1.2\times 10^{-33}\,h_{70}\,^{0.7} 
\beta\;\;\;  {\rm gm\;cm}^{-3}. \\  
 I & = & 16\,h_{70}^{0.7} \;{\rm nwt\;m}^{-2}\,{\rm sterad}^{-1}.  
\end{eqnarray}

\item Helium, carbon etc in white dwarfs and red-giant interiors. 

Here we use eq (6) without the $(1+a)$, since most of the nuclear energy 
is already released on reaching this stage.  Assuming most stars to 
belong to an old population so that 
\begin{eqnarray} 
\rho_{\rm WD}&\simeq &0.1\rho_ * = 3.5\times 10^{33}\,h_{70}\;
\;{\rm gm\;cm}^{-3},\\     
I & = & \frac{0.007c^ 3 \rho_{\rm WD}}{4\pi\langle 1+z\rangle} 
= \langle\frac{3}{1+z}\rangle \,5\times 10^{33}\,\rho_{\rm WD}.\\
& =& 18\,h_{70} \;{\rm nwt\;m}^{-2}\,{\rm sterad}^{-1}. 
\end{eqnarray} 

\item AGN contribution

Madau \& Pozzetti (2000) and Brusa, Comastri \& Vignali (2001) have 
made estimates of the AGN contribution to EBL based on the the abundance 
of massive black holes and that of obscured hard X-ray  
sources, respectively.  They agree that the 
contribution is quite small, of order 5 nwt m$^{-2}$ sterad$^{-1}$.   
\end{itemize}

The upshot is that these readily identifiable contributions add up to 
48  nwt m$^{-2}$ sterad$^{-1}$, well within range (given the obvious 
uncertainties in mean $z$ and other parameters) of the lower estimate 
given at the beginning of this section. It is interesting to note that 
white dwarfs and intergalactic metals come out as the major contributors, 
either one predominating over metallicity in known stars. To reach the higher 
estimate may involve some more stretching of the parameters.  

I thank Richard Bower, Jon Loveday and Max Pettini for helpful 
information and comments.

\end{document}